# Submesoscale circulation in the Northern Gulf of Mexico: Surface processes and the impact of the freshwater river input

Hao Luo[1,2], Annalisa Bracco[1*], Yuley Cardona[1,3], Jim C. McWilliams[4]

[1] School of Earth and Atmospherics Sciences, Georgia Institute of Technology, Atlanta, GA 30332, USA

[2] Department of Marine Sciences, University of Georgia, Athens, GA 30602, USA

[3] Departamento de Geociencias y Medio Ambiente, Universidad Nacional de Colombia, Sede Medellín, Colombia

[4] Department of Atmospheric and Oceanic Sciences, University of California, Los Angeles, CA 90095, USA

Corresponding Author:

Annalisa Bracco, School of Earth and Atmospheric Sciences,

311 Ferst Dr, Atlanta, GA, 30332, USA

email: abracco@gatech.edu

Ph: 404-894-1749

**ABSTRACT**

The processes and instabilities occurring at the ocean surface in the northern Gulf of Mexico are investigated with a regional model at submesoscale-permitting horizontal grid resolution (i.e., HR with dx = 1.6 km) over a three-year period, from January 2010 to December 2012. A mesoscale-resolving, lower resolution run (LR, with dx = 5 km) is also considered for comparison. The HR run is obtained through two-way nesting within the LR run. In HR quantities such local Rossby number, horizontal divergence, vertical velocity, and strain rate are amplified in winter, when the mixed layer is deepest, as found in other basins. In the model configuration considered this amplification occurs in surface waters over the continental slope and off-shore but not over the shelf. Submesoscale structures consist of a mixture of fronts and eddies generated by frontogenesis and mixed layer instabilities, with elevated conversion rates of available potential energy (APE) into eddy kinetic energy (EKE). In all quantities a secondary maximum emerges during the summer season, when the mixed layer depth is shallowest, barely 15-20 m. The secondary peak extends to the coast and is due to the intense lateral density gradients created by the fresh water inflow from the Mississippi River system. Submesoscale structures in summer consist predominately of fronts, as observed in the aftermath of the 2010 *Deepwater Horizon* oil spill, and their secondary circulations are impeded due to the limited depth of the mixed layer. Freshwater river input is key to the submesoscale activity in summer but modulates it also in winter, as shown with a sensitivity run in which the riverine inflow is absent. Implications for transport studies in regions characterized by intense freshwater fluxes and for submesoscale parameterizations are discussed.

# 1 Introduction

Processes occurring at the oceanic submesoscales, between about 100 m and few tens of kilometers horizontally, critically impact transport and mixing in the upper ocean, modify the mixed layer stratification, and dominate the relative dispersion of tracers and floats on comparable scales (Capet et al. 2008b; Zhong and Bracco 2013). The submesoscales are bounded by the geostrophic quasi-two-dimensional mesoscale at larger scales, where the Earth's rotation and the vertical stratification control the dynamics, and by the ageostrophic three-dimensional turbulence at smaller scales, where the effect of planetary rotation is negligible. Submesoscale processes are therefore affected by a weakening of the geostrophic constraint and provide mechanisms to transition energy from the balanced geostrophic mesoscales to the dissipation scales (McWilliams 2008; McWilliams et al. 2001; Molemaker et al. 2005; Molemaker et al. 2010; Muller et al. 2005). Process modeling to resolve these scales has advanced our understanding of those dynamics (Boccaletti et al. 2007; Capet et al. 2008b; Fox-Kemper et al. 2008; Fox-Kemper et al. 2011; Gula et al. 2014; Molemaker et al. 2010; Taylor and Ferrari 2011), while oceanic observations are generally confirmatory but scarce (McWilliams et al. 2009a; D'Asaro et al. 2011; Poje et al. 2014; Shcherbina et al. 2013). In the present work we characterize submesoscale features in the northern Gulf of Mexico (hereafter GoM). This basin hosts important benthic and pelagic fisheries (NOAA, 2012), as well as more than 20,000 natural hydrocarbon seeps (Peccini and MacDonald, 2008). Between April and July 2010, the northern GoM was severely impacted by the *Deepwater Horizon* spill, the largest oil spill in history, that released about 3 x $10^5$ t gas and between 6 and 8 x $10^5$ t oil in the open waters (Joye et al. 2011; McNutt et al. 2012). During the spill six data-assimilating ocean models were used to track and forecast the oil trajectory through virtual particles (Liu et al. 2011). The horizontal resolution of the models was 5 km or coarser, and none of them was able to capture the complexity of the surface circulation in the region between the wellhead and the Mississippi River mouth, as portrayed by SAR images of the surface oil. The images, together with aerial photos, revealed the presence of numerous, spatially coherent, submesoscale frontal structures, several tens of kilometers long and less than a few kilometers wide, that became more prominent from the end of May onward and contributed to the oil transport and convergence in the late spring and early summer of 2010 (e.g., Fig. 3 in Walker et al. 2011). The prevalence of frontal structures in the northern Gulf of Mexico during the summer was further confirmed by the Grand Lagrangian Deployment (GLAD) conducted in August 2012 (Poje et al., 2014). Current modeling (Mensa et al. 2013) and observational (Callies et al. 2015) evidence supports the existence of a seasonal cycle of submesoscale flows with a minimum during the summer season because their energization depends on the mixed layer depth. For a given lateral buoyancy gradient, the shallower is the mixed layer, the less energetic is the submesoscale flow. The surface buoyancy of the northern Gulf of Mexico, however, is subject itself to seasonal and interannual variability due to the presence of large freshwater inputs through the Mississippi-Atchafalaya River systems.

In this work we investigate the origin of the frontal structures observed in the summer of 2010 with a process oriented study using a regional ocean model at submesoscale-permitting resolution (dx = 1.6 km) over a three-year period, from January 2010 to December 2012 with the hypothesis that the freshwater river input into the northern Gulf may force lateral density gradients that in turn fuel frontogenesis also during the summer season, despite the shallow mixed later depth. Comparing the statistical properties of this integration with properties derived from a lower resolution (dx = 5 km) run, we identify when, where, and how fronts and more generally submesoscale dynamics impacts the surface circulation of the northern GoM with a focus on the off-shelf region (i.e. the area characterized by waters deeper than 200 m). The submesoscale field is characterized in terms of its statistical distribution throughout the year, generation mechanisms, and, most importantly, dependence on freshwater fluxes.

Our analysis elucidates some the challenges faced by modelers when trying to predict the trajectories of surface tracers in the northern Gulf of Mexico or more generally any ocean region characterized by substantial freshwater inputs and active submesoscale flows.

**2 Model setup, domain and forcing fields**

We adopt the Regional Oceanic Modeling System (ROMS), a free-surface, terrain-following, hydrostatic, primitive-equation model (Marchesiello et al. 2003), and we implement the Institut de Recherche pour le Dèveloppement (IRD) version of the code, ROMS-AGRIF 2.2 (Debreu et al. 2012). The model domain extends between 97.98ºW - 80.38ºW and 18.02ºN - 31.02ºN (Fig. 1). The horizontal resolution of the grid is 5 km, and the vertical resolution is 70 terrain-following layers with enhanced resolution near the surface (no less than 15 layers in the upper 200 m in the deepest areas) and close to the bottom. We took advantage of the two-way nesting capability of ROMS-AGRIF to introduce a nested (child) grid with horizontal resolution of 1.6 km covering the region between 96.31ºW - 86.93ºW and 25.40ºN - 30.66ºN (Fig. 1). The model bathymetry is derived from ETOPO2 (Sandwell and Smith 1997), is interpolated at 5 km horizontal resolution and transferred to the child grid without modifying the smoothing. In the following we focus on the nested area, comparing results for the parent (LR for low resolution) and child (HR for high resolution) grids. Differences between the two simulations, however, are not limited to the nested region, as shown in Fig. 2, due to the nature of the two-way nesting technique.

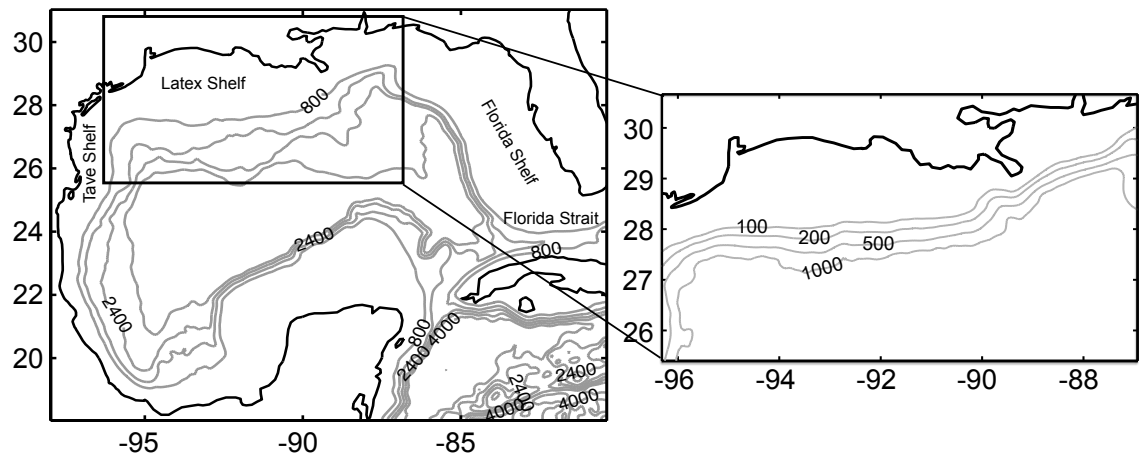

**Figure 1** Model domain and bathymetry. The zoomed region is the nested area where the horizontal resolution is increased to 1.6 km.

ROMS-Agrif is forced by 6-hour surface wind stresses and daily heat fluxes from the European Centre for Medium-Range Weather Forecast ERA-interim reanalysis (Dee et al. 2011; Poli et al. 2010) from December 2009 to December 2012. The resolution of the atmospheric forcing fields is approximately 80 km. At the open ocean boundaries ROMS is nudged to monthly fields derived from HYCOM - NCODA (Hybrid Coordinate Ocean Model - Navy Coupled Ocean Data Assimilation) ocean prediction system (GOMI0.04 expt_30.1) over the period 2009-2012 (http://www7320.nrlssc.navy.mil/hycomGOM). Tidal forcing is generally small in the Gulf of Mexico but for near-shore locations (DiMarco and Reid 1998; Reid and Whitaker 1981), and is neglected. The northern GoM stratification is strongly affected by the freshwater inflow of the Mississippi-Atchafalaya River system, which is generally greatest between May and June and smallest around October. In the key set of integrations discussed in this work only the mean seasonal cycle of the fresh water flux is retained. This is achieved by nudging the surface salinity field to the World Ocean Atlas 2009 (WOA09) monthly climatology (Antonov et al. 2010) with a time scale of 60 days. Further details on this set-up and the validation of the modeled mean circulation are provided in the Appendix. A two-way nested sensitivity run without surface salinity nudging or river forcing is then performed for the year 2010 after an adequate spin-up ($HR_{NOFW}$). The goal of this simulation is to further quantify the role of the freshwater forcing.

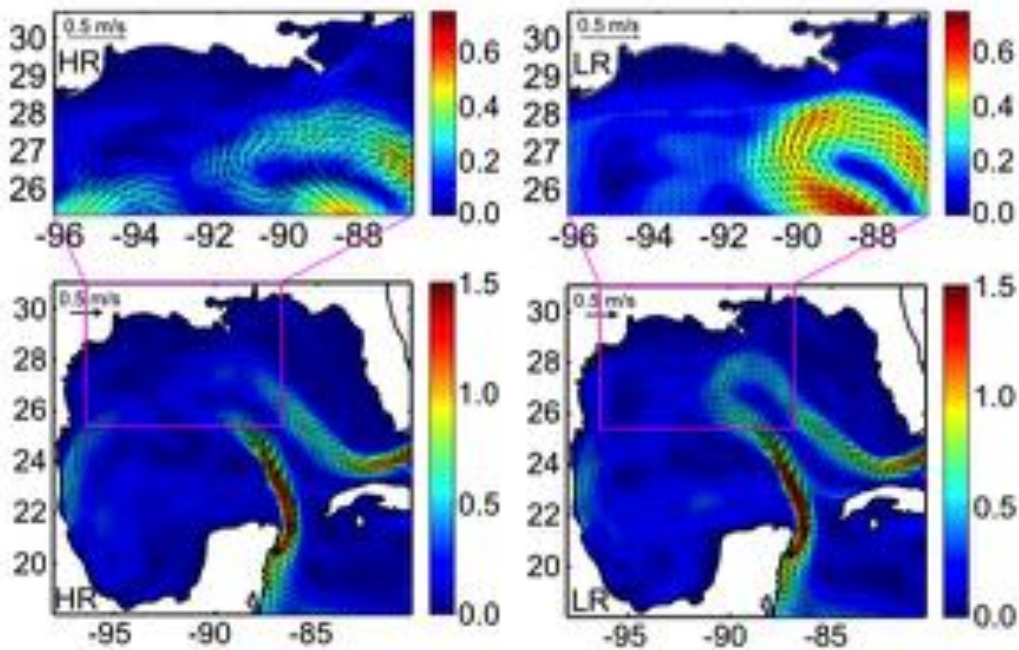

**Figure 2** Modeled surface mean velocities over 2010-2012 superimposed on the mean speed, $\sqrt{\langle u^2 \rangle + \langle v^2 \rangle}$, in LR (left) and HR (right). Insets show the detail of the nested area where the two runs differ in horizontal resolution.

## 3 The annual cycle of surface submesoscale dynamics

In analyzing the distribution of mesoscale and submesoscale features in the northern GoM, four characteristics of the basin should be kept in mind. Firstly, the northern Gulf of Mexico has a wide, shallow continental shelf, defined in the following as the region where the water column is less than 200 m deep, that extends for O(100) km everywhere except in the region between the Mississippi Canyon and the Mississippi River delta; secondly, the continental slope is rather steep to the east of the Mississippi Fan, and broader and more complex to the west; thirdly, the surface layers are influenced by the input of freshwater by the Mississippi-Atchafalaya River system that usually intensifies in late spring and summer; finally the atmospheric circulation is characterized by two distinct seasons with Southeasterlies winds blowing between April and August and stronger Northeasterlies being predominant from September to March. Spring and summer winds can display substantial variability in their directionality particularly to the east of the Mississippi river mouth, and are generally weaker than in fall or winter except for the occasional passage of tropical storms. Fall and winter winds are stronger on average and their intensity varies greatly on time scales of two or three days due to synoptic scale storms.

To isolate the role of freshwater inputs we performed a year-long sensitivity run where no nudging or river inflow are prescribed. The $HR_{NOFW}$ simulation is spun-up for two months from the same December 2009 initial conditions used for the HR case and then run from January to December 2010. Shelf waters shallower than 50 m in proximity of the Atchafalaya and Mississippi mouths retain a portion of the salinity anomaly contained in the initial conditions (Fig. 3). In this section we compare model outcomes from HR, LR and when deemed relevant $HR_{NOFW}$.

We stress that this process study focuses on integrations that retain the seasonal cycle of the freshwater fluxes but not their interannual variability. The resolution adopted, the smoothing of bathymetric features smaller than 5 km, the use of surface salinity nudging to WOA09 data, and the temporal (6-hourly) and spatial (roughly 80 km) resolution of the wind forcing do not allow us to investigate the details of the propagation of freshwater over the shelf, as done for example in Zhang et al. (2012), or the mechanisms responsible for the cross-shelf transport of salinity anomalies to the off-shore waters, discussed for example in Morey et al. (2003). We assume a surface salinity distribution that reflects in its long-term mean the climatologically observed one. In the Appendix we show that over open waters and around the *Deepwater Horizon* site the modeled distributions of summer temperature and salinity are close to observed throughout the water column.

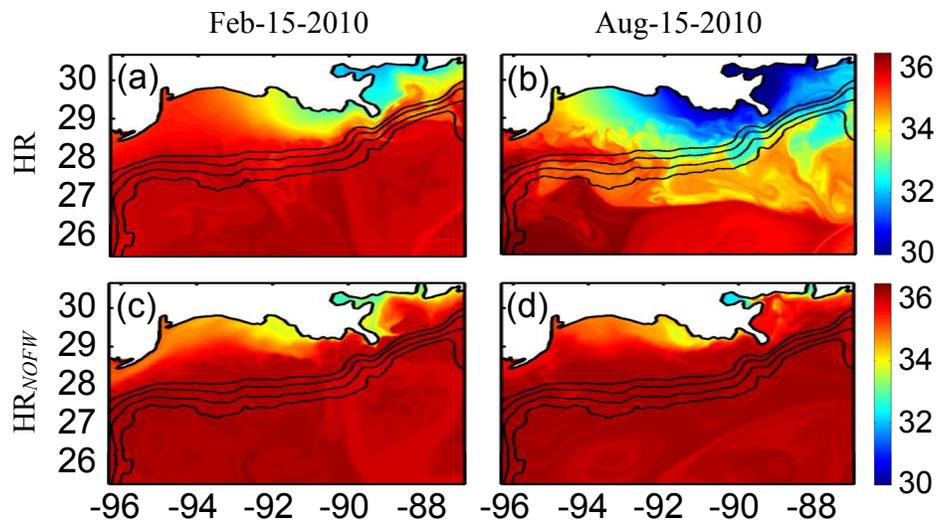

**Figure 3** Snapshots of surface salinity in HR (top) and $HR_{NOFW}$ (bottom) in winter (left) and summer (right) 2010. Unit: PSU

**3.1 Surface vorticity and mixed layer depth**

Submesoscale features are characterized by a local Rossby number $R_o = |\zeta/f|$ of order O(1) and can be identified by their elevated normalized relatively vorticity, $\zeta/f$, in the snapshots of Fig. 4 in mid-February, May, August and November 2011. A large number of submesoscale eddies and fronts are found in HR in all seasons, but their distribution varies greatly through the year.

In winter, between January and March, submesoscale eddies are most numerous, as commonly found in other regions of the world ocean (Callies et al. 2015; Capet et al. 2008a; Capet et al. 2008b; Mensa et al. 2013; Qiu et al. 2014). In this season surface buoyancy gradients and vorticity filaments can intensify and undergo frontogenesis (McWilliams et al. 2009b). Secondary circulations then develop in the vertical in the form of upwelling on the warmer side of the front and downwelling on the colder side in response to the increased strain rate (Capet et al. 2008c, Klein and Lapeyre 2009), and submesoscale mixed layer instabilities contribute to the generation of small eddies by extracting energy from the deep mixed layer (Boccaletti et al. 2007; Fox-Kemper et al. 2008; Molemaker et al. 2005; Thomas et al. 2008). Submesoscale structures form everywhere, including around and within the Loop Current (LC) and its detached Loop eddies, except over the continental shelf, where the mixed layer depth (MLD) reaches the bottom. In LR those instabilities are not resolved, and the relative vorticity is largely underestimated. In spring (April to June) and fall (October to December), fewer submesoscale structures form independently of resolution and differences between HR and LR vorticity fields are less pronounced. Finally in summer (July to September), when the mixed layer is the shallowest, both runs show an increase in the number of submesoscale frontal structures compared to the previous and following seasons. Those structures intrude into the shelf.

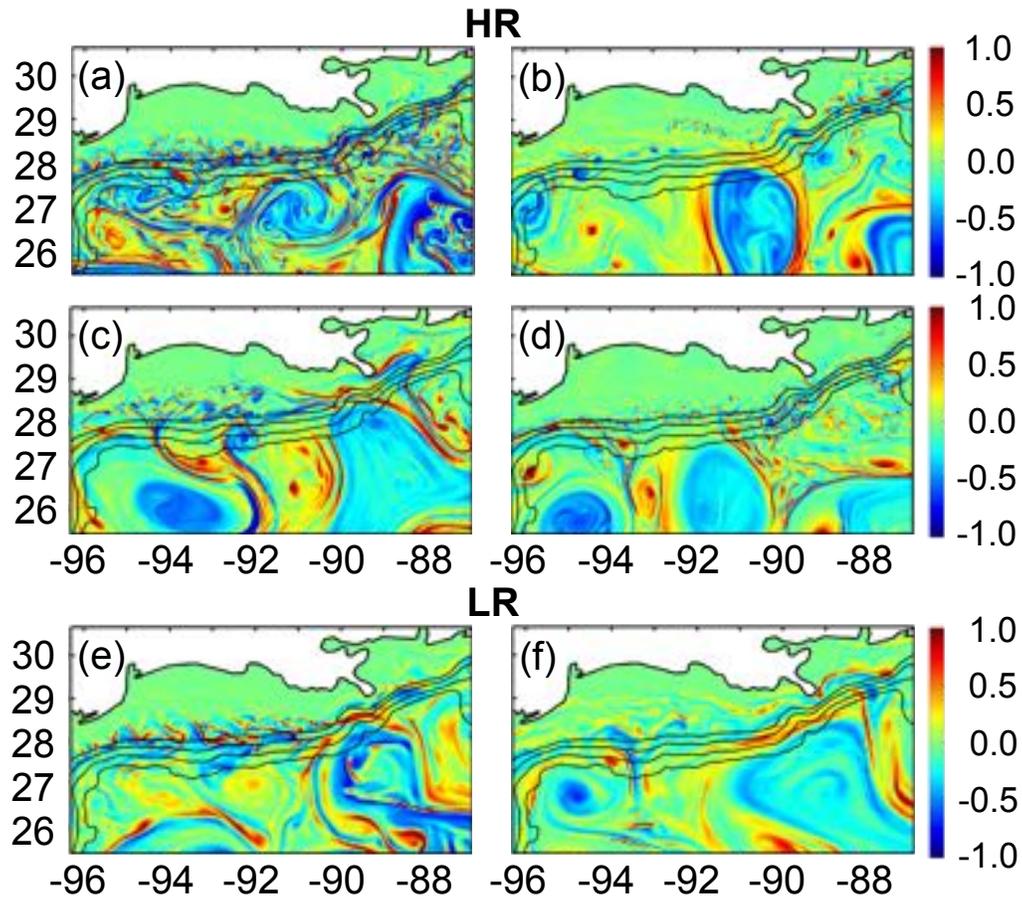

**Figure 4** Snapshots of normalized relative vorticity, $\zeta/f$, in the HR integration on February (Panel (a)), May (Panel (b)), August (Panel (c)), and November (Panel (d)) 15, 2011, and in the LR integration for winter (02/15/2011) and summer (08/15/2011) (Panels (e - f)).

The seasonal variability identified in the maps is confirmed by the analysis of the temporal variability of $R_o$ at the ocean surface, and of the mixed layer depth (MLD) (Fig.

5). Between January and March vorticity is greatest and concentrated offshore waters (> 200 m total depth). Model resolution is key to the representation of the timing of formation (end of December – early January in HR, one month later in LR) and strength (twice as strong in HR) of the submesoscale fronts and eddies that contribute to the winter vorticity peak. In our model, with the limitation of the resolution and bathymetry smoothing adopted, submesoscale structures do not form whenever the mixed layer reaches the bottom and this condition is realized in winter along most of the shelf. In this season the predominant northeasterly winds preclude the transport of submesoscale eddies formed in deeper areas into the shelf that consequently appears by large free from vorticity structures. In summer $|\zeta/f|$ has a secondary maximum, detected in both shallow and deep waters and at both resolutions, despite the average mixed layer reaches its minimum. In this season the area where the MLD reaches the bottom and therefore without strong vertical structures is limited to the portion of the domain shallower than 20-30 m, as shown in Fig. 6. The secondary peak distinguishes the Gulf of Mexico from other ocean regions, such as the North Atlantic (Callies et al. 2015; Mensa et al. 2013), the South American slope (Capet et al. 2008a), or the North Pacific (Qiu et al. 2014 where the intensity and relative importance of submesoscale processes have been found to be inversely proportional to the mixed layer depth. The summer peak dominates the yearly variability of vorticity for waters shallower than 200 m but deeper than the MLD.

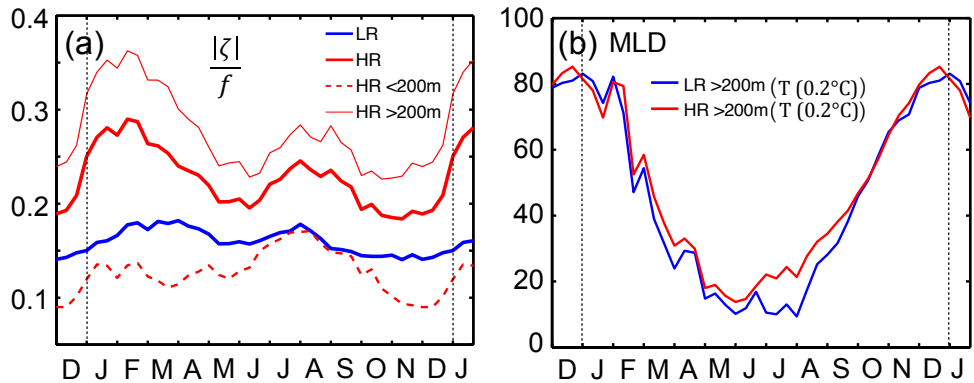

**Figure 5** Annual cycle calculated over three years of (a) the local Rossby number, $R_o = |\zeta/f|$, at the ocean surface averaged over the nested domain in HR (red) and in LR (blue), and (b) of mixed layer depth (MLD) defined using a temperature criterion. The yearly evolution of $R_o$ in HR is shown for the whole region (solid thick lines), over areas where the water column is > 200 m (solid thin lines) and < 200 m (dashed). The averaged MLD is calculated over areas where the water column is > 200 m.

The MLD is defined as the depth at which temperature differences with the surface are equal to 0.2°C, a criterion commonly used in the Gulf. The time series are obtained averaging over grid points where the water column is at least 200 m deep to exclude regions where it reaches the bottom in winter. Very similar curves are obtained using thresholds of 0.1 or 0.3°C, but approximately 4 m shallower or deeper in all months, respectively. The domain averaged MLD is 85 m in both runs in winter. The deepest values are found inside the Loop Current and the Rings, and achieve up to 200 m in few instances, again in agreement with estimates compiled by Weatherly (2004) using nearly 1,500 temperature profiles. Resolving the submesoscale processes causes the modeled MLD to be deeper in summer in all years by approximately 10 m (Fig. 6), and in agreement with observations (Weatherly 2004). We will further discuss this difference in Section 4.

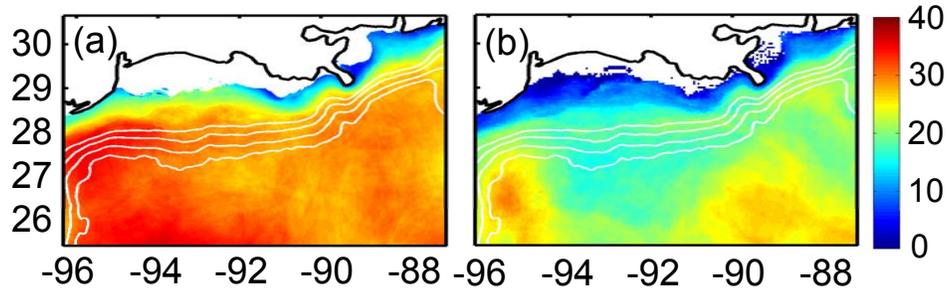

**Figure 6** Mean mixed layer depth calculated using a temperature criterion during the summer season (JAS) in (a) HR and (b) LR. Units: m.

The seasonal probability density functions (PDF) of $|\zeta/f|$, normalized to have unit integral, displays in HR a continuous decrease in the largest positive values attained from winter (JFM) to fall (OND), while negative maxima do not vary significantly throughout the year (Fig. 7). Cyclonic vorticity maxima in HR are linked to vortex stretching associated with frontogenesis at finite Rossby number (Capet et al. 2008c), and are at least twice as large as in LR. Additionally, when the submesoscale dynamics are partially resolved, the distributions maintain a larger positive skewness, $Sk$, in all seasons, with more noticeable differences between winter ($Sk = 2.8$) and spring ($Sk = 2.1$), and the remaining months (summer: $Sk = 1.60$, fall: $Sk = 1.50$). In comparison in the LR case $Sk$ varies only between 1.2 in winter and 0.70 in summer. Noticeably in HR the extremes in vorticity attained in spring are larger than those found in summer (Fig. 7) but the mean vorticity is greater in July-September than April-June (Fig. 4). In this run the cyclonic circulation cells that surround the LC and the Loop eddies are the major contributor to the vorticity extremes. In winter those cells are strongest and deepest than in the rest of year and delimit the MLD local maxima found in the Loop Current and detached eddies, marking also the largest gradients in MLD. They remain stronger than any other structure through April, when the MLD in the LC and Loop eddies remains as deep as 80-100 m. By summer their signature in both vorticity and MLD is comparable to that of the other numerous submesoscale structures.

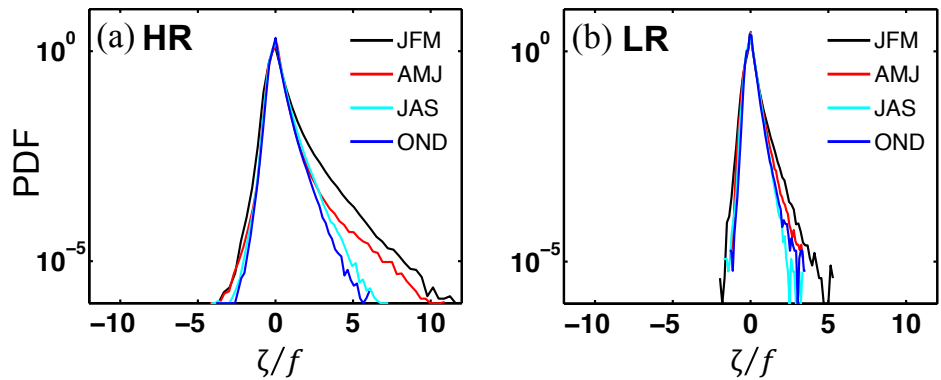

**Figure 7** Seasonal PDFs of | $\zeta/f$ | in (a) HR and (b) LR normalized to have unit integral probability.

In the absence of freshwater inputs the formation of submesoscale fronts and eddies in summer is inhibited (Fig. 8). The seasonality of submesoscale processes in $HR_{NOFW}$ resembles that seen in other basins and the secondary peak in the annual cycle of vorticity is not present (Fig. 9). This difference cannot be attributed to a lower level of mesoscale activity. In this run the mesoscale variability (Loop Current and Rings) measured by the EKE is similar to that of the HR case between January and August 2010 and diverges thereafter with slightly higher EKE levels in early summer and fall due to a different location of the LC in the model domain (Fig. 9). The representation of mixed layer depth is also modified by the river input in both summer and winter. In summer the mean MLD, defined by the temperature criteria, is shallower in the HR$_{NOFW}$ run than in HR in summer and similar to that of the LR case, due to the absence of large vertical velocities associated with submesoscale structures. In winter, especially from late January to the end of February, the ocean surface responds more efficiently to the atmospheric forcing due to the lack of a thin freshwater layer confined at the ocean surface, submesoscale processes intensify, resulting in stronger and slightly more numerous eddies and fronts and therefore higher mean vorticity (Fig. 9), and the MLD deepens slightly over most of the domain (Fig. 10).

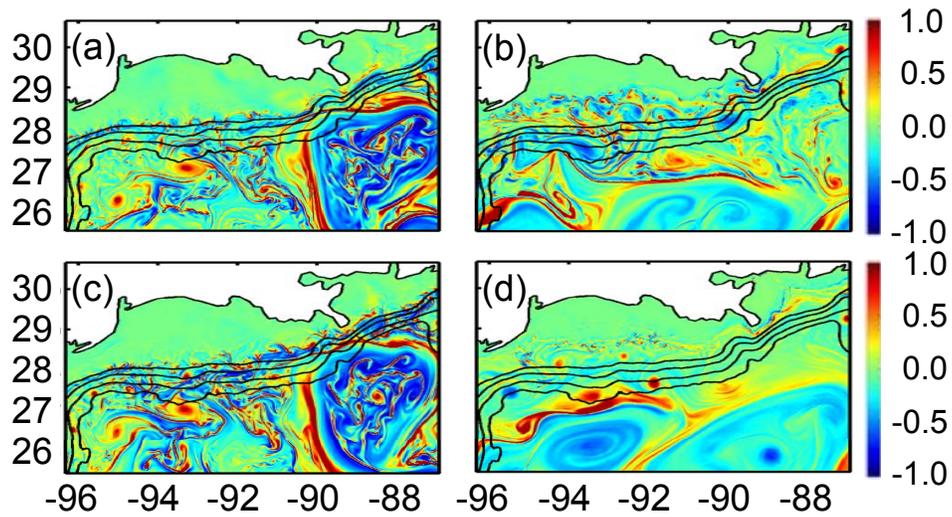

**Figure 8** Snapshots of normalized relative vorticity, |ζ/$f$|, in February (a-c) and August (b-d) 2010 in the HR (left) and HR$_{NOFW}$ (right) runs.

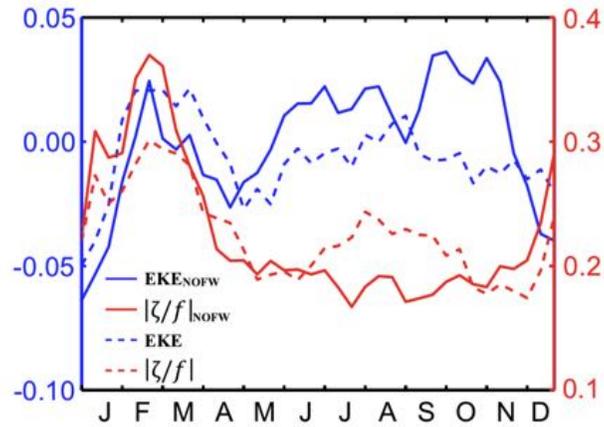

**Figure 9** Time series of |ζ/f| (red) and EKE (blue) from January 1st, 2010 to December 31st, 2010 in the HR$_{NOWF}$ (solid lines) and HR (dashed lines) simulations.

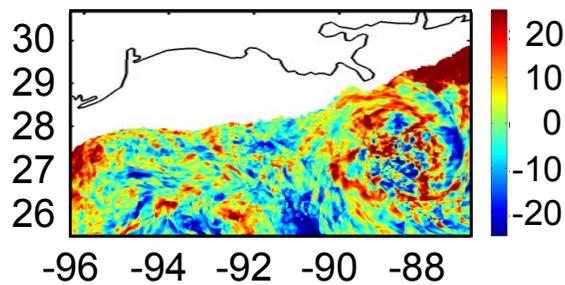

**Figure 10** Difference in mixed layer depth MLD during February 2010 between HR$_{NOFW}$ and HR. Unit: m. Only areas where the water column is > 200 m are shown; the domain average is 6.8 m.

We remark that the resolution adopted in our runs, while reasonably high and comparable to that used in other studies of submesoscale seasonality, is not sufficient to capture the whole range of submesoscale dynamics of the northern GoM. The LR integration captures the mesoscale features and the largest submesoscale eddies, while HR resolves a greater portion of the surface submesoscale features and frontal systems away from the continental shelf. HR, however, does not resolve everywhere with at least three grid points the mixed layer deformation radius $L_{D,ML} = \frac{1}{f}\sqrt{g'H_{ML}}$, where $H_{ML}$ is mixed layer depth defined here using a density criterion as the depth at which density differences with the surface are equal to 0.03 kgm$^{-3}$ and $g'$ is based on the density change at its base. $L_{D,ML}$ can be locally as small as 3.5 km over the broad and shallow shelf in all seasons and limited portions of the off-shore domain in summer (Fig. 11).

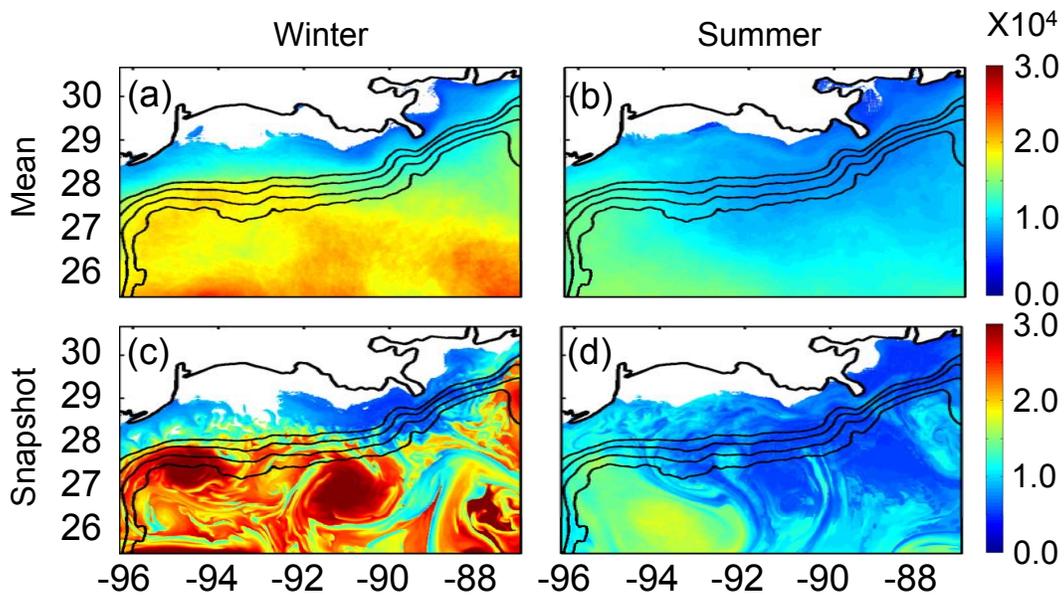

**Figure 11** Top: mean $L_{D,ML}$ calculated for the winter (JFM) and summer (JAS) seasons over three years. Bottom: snapshots of $L_{D,MD}$ in February and August 15, 2011. The minimum value of $L_{D,MD}$ attained in August 15 is 4.7 km and is located to the east of the Mississippi mouth. Regions where the mixed layer reaches the bottom (white in figure) are excluded. Unit: m.

**3.2 Vertical velocities, horizontal velocity divergence, and strain**

The seasonal cycle of the vorticity field is closely followed by the vertical velocity $w$ (Fig. 12, calculated at 4 m depth), by the horizontal velocity divergence $(\partial u/\partial x + \partial v/\partial y)$, that close to the surface is highly correlated to $w$ because of the continuity relation, and by the horizontal strain $S$ defined as $S = \left[\left(\dfrac{\partial u}{\partial x} - \dfrac{\partial v}{\partial v}\right)^2 + \left(\dfrac{\partial v}{\partial x} + \dfrac{\partial u}{\partial y}\right)^2\right]^{1/2}$ (not shown). Again the absolute mean values of all quantities in HR decrease moving from winter into spring, display a secondary maximum in summer, and reach their minima in fall, to grow back steeply in December (Fig. 13). The highest values of w and horizontal velocity divergence are attained where the water column is deeper than 200 m, while the secondary summer peaks interests also the continental shelf. In LR the winter peak is decisively underestimated and only the fall minimum can be separated from the other months.

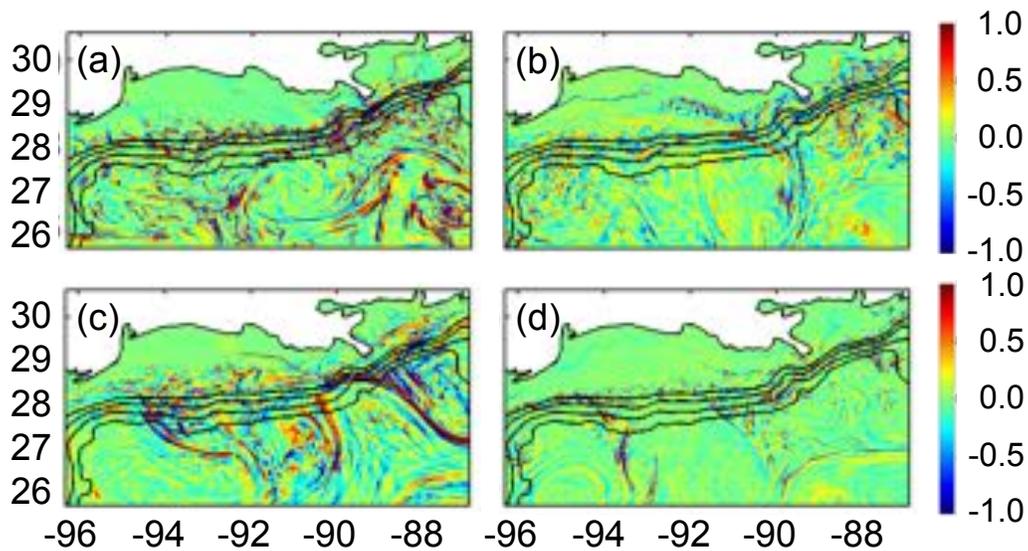

**Figure 12** Snapshots of vertical velocity $w$ calculated at 4 m depth in the HR integration on February (Panel (a)), May (Panel (b)), August (Panel (c)), and November (Panel (d)) 15, 2011. Unit: $10^{-3}$ m s$^{-1}$.

The seasonal PDFs for vertical velocity and horizontal divergence are similar in shape to the vorticity ones but with opposite asymmetry around the y-axis (not shown). Again, in HR the largest values are achieved in winter followed by spring and are associated with the circulation cells around the Loop structures that extend to several hundred meters into the water column. The LR integration underestimates the tails of the distributions for both quantities by at least a factor of 2 when compared to HR.

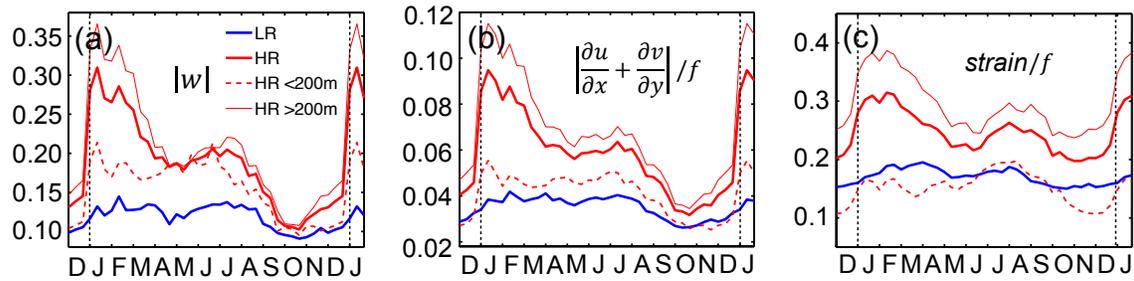

**Figure 13** Annual cycle calculated at 4 m depth and using the three year long simulation of (a) $|w|$, (b) $\left|\left(\frac{\partial u}{\partial x}+\frac{\partial v}{\partial y}\right)/f\right|$, and (c) strain $S/f$ (see text) averaged over the nested domain in HR (red) and in LR (blue) in the whole nested region (thick solid line), in HR over areas where the water column is < 200 m (dashed), and > 200 m (thin solid). Unit of $w$: $10^{-3}$ m s$^{-1}$.

In the absence of river forcing the seasonal peaks in near surface vertical velocity, horizontal divergence and strain are suppressed in summer and slightly amplified in winter, analogously to what shown for vorticity.

**3.2 River inflow and density gradients**

Submesoscales are apparent as lateral (horizontal) density gradients and intense circulations throughout the ocean surface. Those gradients are strained and stirred by the mesoscale eddies, and are usually associated with frontogenesis and ageostrophic circulations. In the northern Gulf of Mexico, as in other basin impacted by large river outflows, density gradients are not driven only by the mesoscale circulation but also by the input of freshwater from the Mississippi River system. In the Gulf such input reaches its maximum in May-June and spreads into the northern portion of the basin during summer (Gierach et al. 2013). Fresh water anomalies are transported over the broad, shallow continental shelf by the wind-driven along-shelf currents (Cardona and Bracco 2014; Zavala-Hidalgo et al. 2003). The offshore, cross-shelf transport, on the other hand, is likely contributed by the numerous eddies that form over the continental slope and impinge on the shelf preferentially between April and August (Marta-Almeida et al. 2013; Ohlmann and Niiler 2005), and in several instances by the Loop Current that extends sufficiently far north to entrain the freshwater at its periphery.

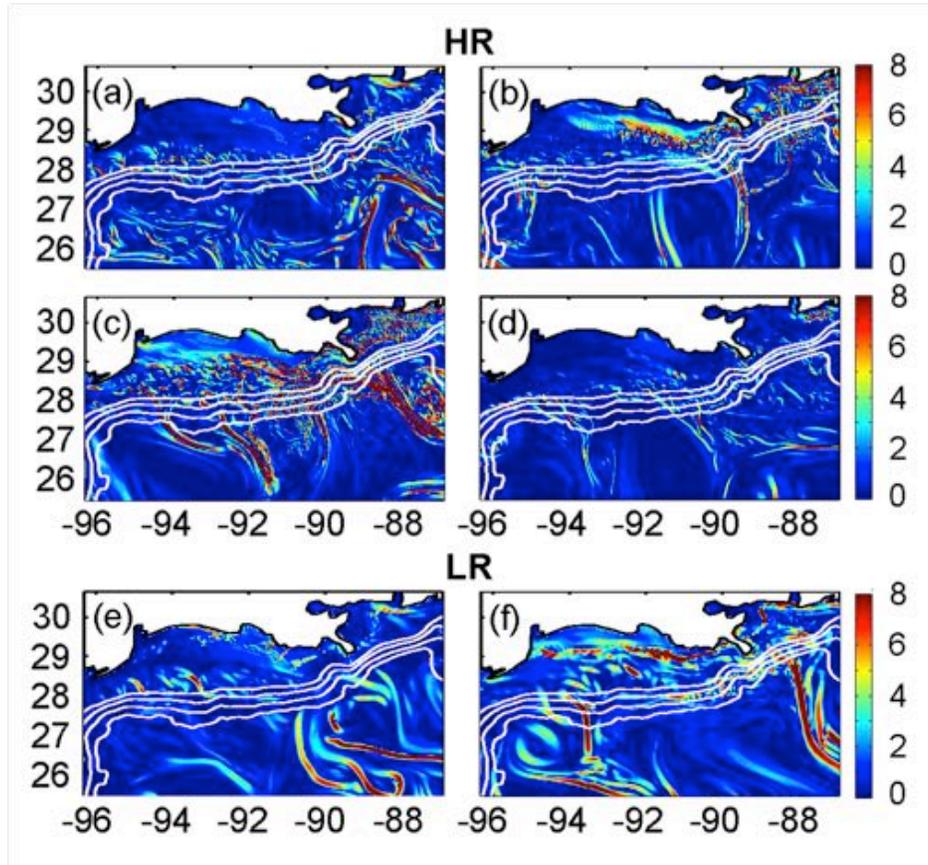

**Figure 14** Snapshots of surface horizontal density gradients, $|\nabla_h \rho|$, in the HR integration in February (Panel (a)), May (Panel (b)), August (Panel (c)), and November (Panel (d)) 15, 2011, and in the LR integration for winter (02/15/2011) and summer (08/15/2011) (Panels (e - f)). Unit: $10^{-5}$ kg m$^{-4}$.

The seasonal evolution of $|\nabla_h \rho|$ is exemplified in Fig. 14 in HR and LR and time-series are presented for all three runs analyzed in this work in Fig. 15. When the riverine input is included surface density gradients are greater in summer than in any other season, followed by spring, winter and fall, and an increase (decrease) in the mean gradient amplitude coincides with larger (smaller) extreme values (not shown). The seasonal cycle of $|\nabla_h \rho|$ follows with one-to-two months delay the climatology of the river input into the ocean that increases from winter to late spring and decreases from the end of June into fall, with a minimum in early October. The largest gradients are found over the shelf, where they are approximately 40% larger than for offshore waters, with maxima near the mouths of the major two rivers, Mississippi and Atchafalaya; in open ocean waters, on the other hand, enhanced values are localized around the mesoscale Loop Current and the Rings in summer and both around and within those large mesoscale structures in winter. Finally, over the shelf differences between HR and LR are limited to smaller amplitude in the lower resolution case, and the relative differences between seasons are reproduced independently of the grid spacing, while offshore the dominance of the summer peak is lost at 5 km resolution. In the absence of freshwater input the offshore density gradients are amplified by the mesoscale field in winter and are greatly reduced in summer, following closely the MLD evolution. Over the shelf the salinity anomalies from the initial conditions are not fully removed or homogenized in $HR_{NOFW}$ and maintain an approximately constant value through the integration.

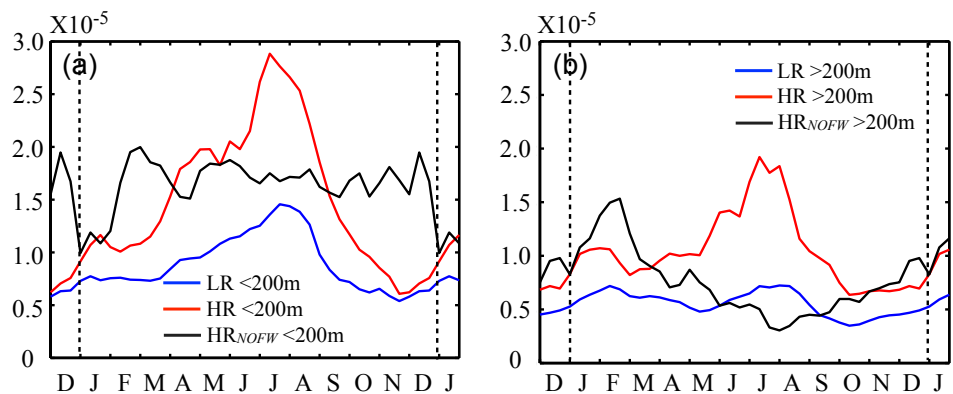

**Figure 15** Annual cycle of $|\nabla_h \rho|$ averaged areas where the water column is (a) < 200 m and (b) > 200 m over the nested domain in HR (red), LR (blue) and $HR_{NOFW}$ (black). Unit: kg m$^{-4}$.

**4 Interpretation of the submesoscale variability in the Gulf of Mexico**

We have shown that in the northern GoM submesoscale processes affect both the distribution and the seasonality of all quantities, from vorticity to lateral density gradients, and that in presence of freshwater forcing high level of submesoscale activity are supported also during the summer season. We now investigate the submesoscale dynamics in more detail focusing on the HR and $HR_{NOFW}$ solutions.

Time series, maps and PDFs of vorticity and other variables suggest that during the winter season the submesoscales are populated by fronts and eddies, as found in other regions, whose generation commonly result from frontogenesis, mixed layer instabilities destabilizing the fronts (MLIs) (Boccaletti et al. 2007; Capet et al. 2008b; Haine and Marshall 1998; Mensa et al. 2013; Thomas and Ferrari 2008; Thomas et al. 2008) and possibly symmetric instabilities inside the large anticyclonic Rings (Brannigan et al., 2015). In summer, on the other hand, the vorticity fields are populated by fewer submesoscale eddies and a large number of frontal structures.

In winter, when the MLD is deep, the frontal structures within the mixed layer are prone to secondary circulations (Capet et al. 2008d) and MLIs (Boccaletti et al. 2007) with the generation of submesoscale eddies; in summer, when the mixed layer is shallow and stores less APE, the growth of restratifying baroclinic instabilities along the front edges is inhibited (Mensa et al., 2013; Callies et al., 2015). In idealized and realistic model configurations where mesoscale fronts and eddies dominate the dynamics, frontogenesis has been shown to increase for increasing mixed layer depth (Badin et al. 2011; Fox-Kemper et al. 2008; Levy et al. 2011; McWilliams et al. 2009a; McWilliams et al. 2009b). The strain field at the edges of eddies contributes to sharpen the surface buoyancy gradients with the generation of fronts that can eventually become unstable, and those gradients are generally more energetic the deeper is the MLD. Brannigan et al. (2015), on the other hand, used a suite of simulations in which the strength of the winds was kept constant through the year and only heat fluxes underwent a seasonal cycle to show that decoupling the wind strength from the heat fluxes allows for an intensification of frontogenetical processes when the mixed layer is shallow. Additionally, they experimented with resolution varying from 4 to 0.5 km and found that for increasing resolution all submesoscale structures become stronger and more abundant, MLI and symmetric instabilities occur more frequently year around, but the degree of inhibition of MLIs and symmetrical instabilities in summer compared to winter is independent of resolution once the submesoscale processes are partially resolved.

In the northern GoM density gradients are not controlled only by the mesoscale or large-scale flow and the atmospheric forcing, but depend on the interplay between those two factors and the freshwater fluxes introduced by the Mississippi River system. As a result, density gradients are abundant in summer and winter around the LC and the Rings and between the mesoscale structures in offshore waters, and they occupy more prominently than in other seasons the shelf during summer and the interior of the LC and Rings in winter.

To quantify the relative role of fronts and MLIs in the GoM, first we consider the frontal tendency at the ocean surface, then we separate the mesoscale and submesoscale fractions of the flow following Mensa et al. (2013) and we investigate the energy transfers calculating the conversion rate of available potential energy (APE) to eddy kinetic energy (Capet et al. 2008c; Mensa et al. 2013).

The flow frontal tendency, $F$, is defined as $F = \dfrac{D|\nabla_h \rho|}{Dt} = Q \cdot \nabla_h \rho$ with $Q = (Q_1, Q_2) = -\left( \dfrac{\partial u}{\partial x}\dfrac{\partial \rho}{\partial x} + \dfrac{\partial v}{\partial x}\dfrac{\partial \rho}{\partial y}, \dfrac{\partial u}{\partial y}\dfrac{\partial \rho}{\partial x} + \dfrac{\partial v}{\partial y}\dfrac{\partial \rho}{\partial y} \right)$ (Capet et al. 2008c; Hoskins 1982; Hoskins and Bretherton 1972). Fig. 16 presents maps of $F$ in February and August 2010 in HR and $HR_{NOFW}$, and Fig. 17 follows with the time series of the mean monthly values averaged over the nested domain over three years in the HR case and 2010 for $HR_{NOFW}$. A positive frontal tendency indicates an increase of the magnitude of the density gradient over time and therefore frontogenesis, while a negative sign implies frontolysis. In HR $F$ peaks in summer, at the time when lateral density gradients are greatest, and displays a secondary maximum in winter, when the mixed layer is deepest, with intermediate values in spring and minima in fall. In $HR_{NOFW}$ the winter peak is almost twice as strong due the more efficient response to the atmospheric forcing, the summer maximum is absent, and late fall is characterized by intermediate values.

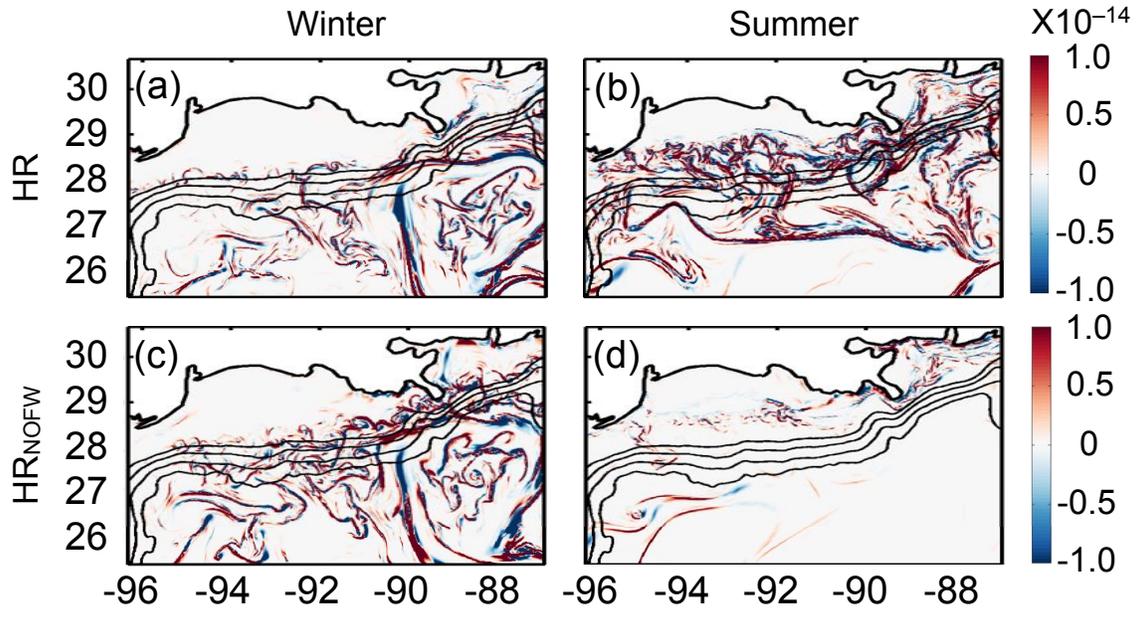

**Figure 16** Snapshots of frontal tendency $F$ in (a) February, and (b) August 2010 in HR (top) and $HR_{NOFW}$ (bottom). Unit: $10^{-14}$ kg$^2$ m$^{-8}$ s$^{-1}$.

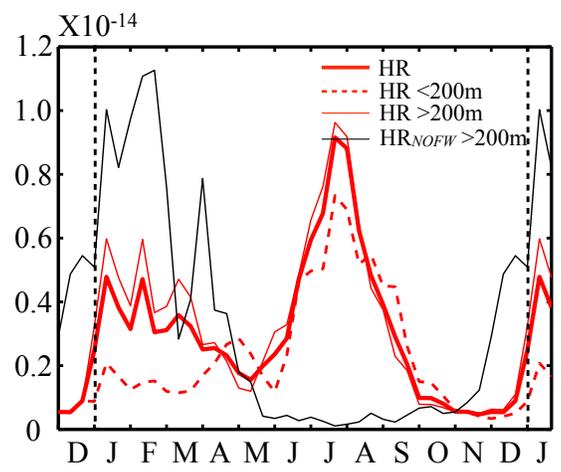

**Figure 17** Annual cycle of frontal tendency $F$ in HR averaged over the nested domain and over three years in the whole region (thick solid line), over areas where the water column is < 200 m (dashed), and > 200 m (thin solid) and in $HR_{NOFW}$ over areas where the water column is > 200 m (black solid line). Unit: kg$^2$ m$^{-8}$ s$^{-1}$

The model representation of frontogenesis affects that of MLD. In summer the mixed layer depth is deeper in HR than $HR_{NOFW}$ and LR due to the increased vertical mixing induced by the numerous fronts that erode its shallow base. Deepening of the MLD due to submesoscale frontogenesis in regions of warm and shallow-mixed layers has been found also in the subtropical gyre, but for winter months (Lévy et al. 2010). In winter in HR the simultaneous presence of submesoscale fronts but also of restratifying submesoscale eddies produces compensating effects and the mixed layer depth is similar to that of LR, while in $HR_{NOFW}$ the absence of a freshwater layer causes a small deepening.

Overall frontogenesis follows closely the MLD evolution in $HR_{NOFW}$ and the annual cycle of $|\nabla_h \rho|$ in HR, despite similar levels of mesoscale activity. In winter F displays comparable values in HR and $HR_{NOFW}$ inside the LC and Rings, being the core of those mesoscale structures isolated from the freshwater input, while is higher in $HR_{NOFW}$ between those transport barriers. In summer F is negligible inside the Rings or the LC extension in both cases, but is strong around them and everywhere the water column is deeper than 30 m in HR.

We then quantify the APE to EKE conversion rate, *PK*, associated with the submesoscale dynamics. *PK* is defined as $PK = \frac{1}{MLD} \int_0^{-MLD} \langle w'b' \rangle_{xy} dz$, where $\langle \ \rangle_{xy}$ indicates the area average over the model domain, again excluding areas where the mixed layer reaches the bottom (Boccaletti et al. 2007; Capet et al. 2008d; Fox-Kemper et al. 2008; Mensa et al. 2013), where *b* is buoyancy defined as $b=-g\rho/\rho_0$, $\rho_0$ is the reference density obtained by averaging over the mixed layer of the nested region, the ' indicates the submesoscale residual. To decompose the flow into its mesoscale component and submesoscale residual we adopt a rotationally symmetric two-dimensional Gaussian low-pass filter with a kernel size of 151x151 grid points and a filtering scale of 60 km. The filtering scale has been chosen based on tests performed on various fields, from vorticity to currents and strain, using scales from 20 to 120 km. Sensible separation and little dependency on the filtering scale was found for values comprised between 50 and 75 km.

In Fig. 18 *PK* is shown together with $\langle |\nabla \bar{b}|\rangle^2_{xyz} \cdot \langle MLD \rangle^2_{xy}$ to test the scaling argument on potential energy conversion by MLIs that is at the base of the parameterization proposed by Fox-Kemper et al. (2008). Here $\bar{b}$ is the mesoscale component of *b* and $\langle \ \rangle_{xyz}$ indicates volume averaging over the HR region and the mixed layer. In this scaling *PK* is considered proportional to $\frac{1}{f}\langle |\nabla \bar{b}|\rangle^2_{xyz} \cdot \langle MLD \rangle^2_{xy}$ under the assumptions that energy conversion by MLIs is proportional to the buoyancy gradients sharpened by the mesoscale strain field, and that such sharpening occurs more effectively the deeper is the mixed layer. A necessary but not sufficient condition for the relationship above to be verified is that $\nabla \bar{b}$ and $\nabla b'$ must follow the same seasonal cycling. In HR and HR$_{NOFW}$ this is verified offshore (Fig. 19) but not over the shelf where, at the resolution considered, the mesoscale field is in essence nonexistent. In calculating the curves in Fig. 18 we therefore excluded the shelf region.

Successful verification of the *PK* scaling in realistic model configurations have been performed by Capet et al. (2008a) for the Argentinian shelf and by Mensa et al. (2013) for the Gulf Stream region. In those domains MLIs have been found to occur in presence of a combination of deep MLD and mesoscale-induced horizontal density gradients, and the two curves follow approximately the same seasonal cycling and can be superimposed with an opportune scaling coefficient. In the GoM this is verified only in $HR_{NOFW}$ (Fig. 18b). In the HR run, on the other hand, *PK* is largest in January and February, sharply decreases in March reaching its minimum at the end of May, increases slightly in mid-June, oscillates around a low mean value through the summer and part of fall, and increases sharply in December. $\langle |\nabla \bar{b}|\rangle^2_{xyz} \cdot \langle MLD \rangle^2_{xy}$ follows closely the *PK* cycling only from mid March to the end of May. During the remaining of the year it grows steadily from June into mid February overestimating during summer and fall and underestimating in winter the PK curve by approximately a factor of two. The divergent behavior is due to the freshwater fluxes that force a seasonal cycling of the lateral density gradients that is opposite to that of the mixed layer. The riverine input is responsible in summer for large values of $|\nabla \bar{b}|$ while MLIs are inhibited by the shallow MLD in summer, in fall for small $|\nabla \bar{b}|$ values despite the rapidly increasing MLD, and in winter for limiting $|\nabla \bar{b}|$ by increasing the near surface stratification.

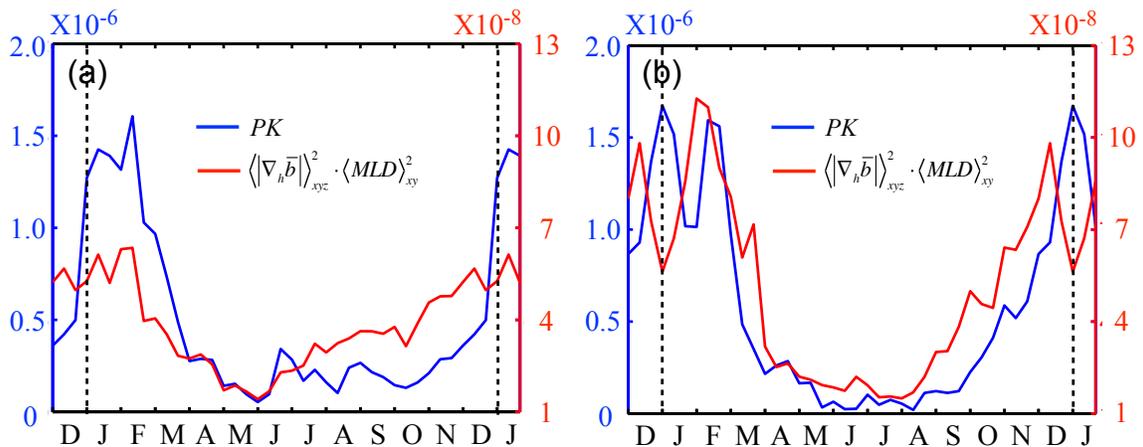

**Figure 18** Annual cycle averaged over (a) 2010-2012 in HR and (b) 2010 in $HR_{NOFW}$ and over the nested domain of *PK* (blue) and of $\langle |\nabla \bar{b}|\rangle^2_{xyz} \cdot \langle MLD \rangle^2_{xy}$. Units: $m^2 s^{-3}$ for *PK* and $m^2 s^{-4}$ for $\langle |\nabla \bar{b}|\rangle^2_{xyz} \cdot \langle MLD \rangle^2_{xy}$.

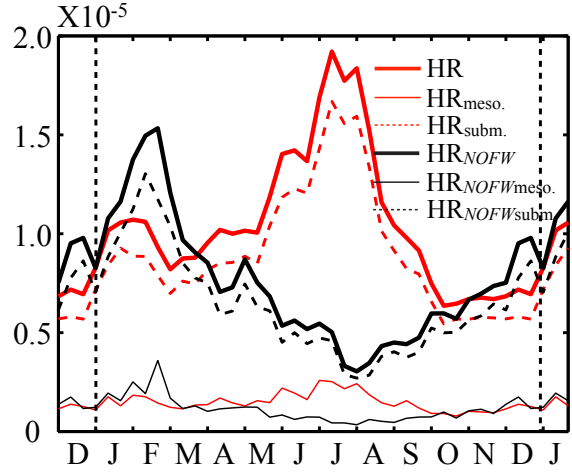

**Figure 19** Annual cycle of $|\nabla_h \rho|$ in offshore waters separated in its mesoscale ($|\nabla_h \bar{\rho}|$, thin solid line) and submesoscale ($|\nabla_h \rho'|$, dashed line) components in HR averaged over three years and $HR_{NOFW}$ for 2010.

## 5. Conclusions

This work characterizes the submesoscale dynamics near the ocean surface in the northern Gulf of Mexico in the presence of a large freshwater input with a series of regional simulations. Two runs account for the surface salinity variations associated with the riverine input from the Mississippi and Atchafalaya River system, cover the 2010-2012 period, and their horizontal resolution varies from dx = 5 km (LR) to 1.6 km (HR), with which the submesoscale processes are either mostly unresolved and partially resolved, respectively. A third run, $HR_{NOFW}$, spans 2010 at 1.6 km resolution without freshwater forcing. In the HR simulation a combination of frontal structures and submesoscale eddies generated by baroclinic instability along the fronts emerge in the winter season, between December and March, when surface mixed layer and the available potential energy are greatest, in qualitative agreement with previous studies in other ocean basins. These eddies and fronts are found predominately offshore of the shelf areas, around and within the large mesoscale structures that populate the Gulf, and are significantly underestimated in LR. Contrary to other geographical areas, however, submesoscale dynamics actively contribute to the generation of a secondary seasonal peak in vorticity, strain rate, horizontal velocity divergence, and vertical velocity in HR in summer, in correspondence with the maximum of river inflow and surface lateral density gradients. We have shown that these density gradients, created by the intense freshwater fluxes in late spring and early summer, feed summer frontogenesis, and that frontogenetic processes are more prominent in summer than in any other season both along the shelf and in offshore waters. Summer fronts, while more abundant and intense than at any other time of the year, develop only weak secondary circulations, and are less prone to mixed layer baroclinic instabilities due to the limited vertical extension of the mixed layer. Summer frontogenesis is associated to a deepening of mixing of the surface warm temperatures and as a consequence, the mixed layer is deeper in the integration at finer resolution by approximately ten meters, a 50% increase compared to the 5 km run. This agrees well with *in situ* observations. The integration without freshwater forcing, on the other hand, does not develop any substantial submesoscale structure in summer, in agreement with previous studies in regions not affected by riverine input, despite a comparable mesoscale field, and is characterized by higher mean vorticity and associated fields and greater conversion of APE into EKE through mixed layer instabilities in winter.

We verified that the scaling motivating the parameterization for mixed layer instabilities proposed by Fox-Kemper et al. (2008) is verified in $HR_{NOFW}$ but does not hold both offshore and over the shelf whenever the freshwater fluxes are included. In presence of freshwater forcing the submesoscale dynamics depends on the atmospheric fluxes, through their first order control of mixed layer depth, on the mesoscale activity of the affected region, but also on the riverine input through their control of frontogenetic processes. This points to the importance of representing correctly surface salinity anomalies in forecast or hindcast applications, and consequently of having a reliable network of salinity measurements in coastal areas, especially in proximity of river inflows.

This study is limited in several respects. The representation of the riverine input through nudging to a monthly climatology does not account for its large interannual variability; the daily run-off of major rivers in the basin should be implemented as next step. We suggest that both the oceanic model resolution and the temporal and spatial resolution of the wind products used to force the ocean model (6-hourly and roughly 80 km in the current configuration) have to be further refined, particularly with reference to the representation of the shelf dynamics. The model resolution in this work precludes us from making definitive statements on the relative roles of frontogenesis versus mixed layer instability statistics in summer because the deformation radius of the mixed layer is barely resolved; it is not even clear that this distinction is well founded because many submesoscale fields exhibit both processes simultaneously. Nonetheless, preliminary simulations at 0.5 km resolutions performed for a comparable domain (Roy Barkan, personal communications), floats trajectories during the GLAD experiment, distribution of *Sargassum* observed in satellite images, and of the surface oil during the summer months following the 2010 spill, all support a predominance of frontal structures at the submesoscales between June and August in the northern Gulf of Mexico, in agreement with the results shown.

Satellite and aerial images of the oil that reached the surface during the 2010 *Deepwater Horizon* oil spill highlighted the unexpected presence of numerous frontal structures where the oil accumulated. Oil-filled fronts emerged at scales ranging from tens of meters (Langmuir scales) to several tens of kilometers (the upper end of the submesoscale range) and contributed to both its shoreward transport and mixing within the euphotic layer. In this work we have focused on the processes behind the development of the submesoscale fronts observed so numerous in the off-shore waters of the GoM in the 2010 summer, but missed by all forecast models run to predict the fate of the oil at the surface. Langmuir lines and Stokes drift effects (McWilliams and FoxKemper 2013; Sullivan and McWilliams 2010) were also critical to the transport of oil away from the wellhead. Open questions concern the structure of the velocity field at the scales from tens of meters to few kilometers (Poje et al. 2014). Simulations investigating the interactions between submesoscale structures (eddies and fronts), Langmuir cells, and wave motions in the Gulf of Mexico environment at different times of the year should be performed to improve any future forecasting effort. Those studies will need to account for the substantial variability of the submesoscale flow in the Gulf of Mexico throughout the year and between different years, as a function of the varying freshwater river input, to characterize and possibly predict the pathways and concentrations of surface material, from pollutants to nutrients, chlorophyll, algae, and nitrogen-fixing cyanobacteria (Cardona et al. 2015; Gower and King 2011; Toner et al. 2003; Zhong et al. 2012) that regularly form large blooms in the offshore and coastal waters of the Gulf. We plan to explore further this question, as well as the relation between submesoscale structures, Langmuir cells, and surface waves with simulation at higher resolution in the near future.

Finally, we speculate that the results presented on the interplay between freshwater fluxes, surface frontogenesis, vertical mixing, and mixed layer depth may be relevant to other oceanic regions where river inputs and/or precipitation or ice-freezing and melting are major contributors to the near-surface density field, and that coastal areas in proximity of riverine estuaries are likely 'hot-spots' of submesoscale activity year-around.

**Acknowledgment** This work was made possible (in part) by a grant from BP/the Gulf of Mexico Research Initiative to support the consortium research entitled "Ecosystem Impacts of Oil and Gas Inputs to the Gulf (ECOGIG)". JM was supported by the Office of Naval Research (N00014-12-1-0939). AB thanks the support provided by the NSF sponsored Institute for Pure & Applied Mathematics at UCLA, where the first draft of this paper was completed. The simulations described here have been (partially) archived at the Gulf Research Initiative Information and Data Cooperative (GRIIDC), database # R1.x132.141:0005. All simulated fields will be made available upon request to the corresponding author.

**Appendix. Model mean circulation and validation**

In the configuration used ROMS-Agrif uses a third-order, upstream-biased scheme for advection, and a Laplacian horizontal diffusion and a rotated split-upstream advection-diffusion scheme (Lemarie et al. 2012; Marchesiello et al. 2009) for momentum and for tracer (temperature and salinity) mixing. A non-local closure scheme based on the K-profile planetary (KPP) boundary layer scheme (Large et al. 1994) is used at the surface boundary layer, and it parameterizes vertical mixing due to shear instability in the ocean interior. The model bathymetry, derived from ETOPO2 (Sandwell and Smith 1997), is smoothed with a Shapiro smoother (Penven et al. 2008) with a maximum slope parameter (i.e. the ratio of the maximum difference between adjacent grid cell depths and the mean depth at that point) $r_{max} = 0.35$ to minimize potential pressure gradient errors. The bathymetry is interpolated at 5 km horizontal resolution and transferred to the child grid for the nested runs. This ensures that differences in the flow dynamics between parent and child grids are associated with the fluid-dynamical resolution and to the model's ability to represent different processes at different scales, rather than to increased complexity of the bottom relief. Independently of the fresh water treatment the low resolution runs are initialized with the HYCOM GOMl0.04 expt_30.1 ocean state on January 1st, 2009, are spun-up for three years repeating the forcing and boundary conditions of 2009, and are continued until the end of 2012. The nested simulations are the initialized on December 1$^{st}$ 2009 at the end of the third year, and run through 2012. In the analysis we consider the period January 2010 to December 2012. The large-scale circulation of the Gulf of Mexico can be approximated by a two-layer system. The upper layer, extending to depths of 800-1200 m, is dominated in the east by the Loop Current (LC) with its shedding of large anticyclones, so-called LC eddies or Rings with typical diameters of 200-400 km, and in the west by the translation of these Rings across the Gulf basin (Hamilton et al. 1999; Vukovich 2007; Welsh and Inoue 2000). The Rings detach from the LC at an irregular frequency of 9-14 months (Sturges and Kenyon 2008; Sturges and Leben 2000; Vukovich 1995) and occupy the upper 800-1000 m of the water column (Cooper et al. 1990; Forristall et al. 1992; Lee and Mellor 2003). Along the shelves, particularly along the Tamaulipas-Veracruz and Louisiana-Texas shelves, the circulation is predominantly wind-driven (DiMarco et al. 2005; Zavala-Hidalgo et al. 2003). From April to August southeasterly winds prevail, while from September to March northeasterlies characterize the downwelling season (Marta-Almeida et al. 2013). The circulation along the Florida shelf, while wind-driven, does not display any clear seasonality (Ohlmann and Niiler 2005). The model reproduces well the features described above. A detailed comparison with in-situ and satellite altimeter data for a model version differing from the one used here in vertical resolution (35 instead of 70 layers), frequency of the atmospheric forcing (monthly instead of six-hourly), and simulated period (2000-2008 instead of 2010-2012) is presented in Cardona and Bracco (2014), and it is not repeated except for the two validations presented below. The configuration adopted here simulates analogously the large scale flow (e.g. Fig. A1) while providing a better matching with in-situ data locally (e.g. Fig. A2), owing to the inclusion of higher frequencies in the forcing fields and to the enhanced vertical resolution. The seasonal cycle and intensity of the surface geostrophic velocities, the statistics associated with the variability and eddy shedding of the Loop Current, and the deep circulation are modeled accurately. The shedding of the Rings is influenced by the transport into the Gulf through the Yucatan Channel (Cardona and Bracco 2014; Chang and Oey 2010a) and by the winds (Chang and Oey 2010b), but it is not a deterministic process (Cardona and Bracco 2014, Lugo-Fernandez 2007, Sturges and Leben 2000). The runs presented here reproduce well the statistics of the observed fields, but the shedding events differ in LR and HR. In particular, the LC in LR is characterized by fewer shedding episodes and therefore less variability than HR over the three years considered. Consequently, the modeled mean velocities at the surface (see Fig. 1) appears slightly higher in HR. This is not due to model resolution, but to the non-deterministic nature of the Ring detachment process that takes place on time scales of 9-14 months (i.e. on scales of the same order of magnitude of the integration length). Indeed the mean circulation in a 8-year long simulation of the whole GoM at 5 km horizontal resolution (ITD3 in Cardona and Bracco 2014) has mean speed patterns indistinguishable from HR (not shown). Figure A1 shows the modeled EKE time series, calculated over the nested area using the velocity anomalies at each grid point and subtracting the total mean calculated over three years. No obvious seasonality can be detected in all three data sets over such a short (compared to the LC shedding) time frame, and the seasonal cycle has not been removed. Given the nature of the LC behavior and the absence of any data assimilation in our runs, only agreement in the mean level and variance of EKE can be expected. The modeled and AVISO-derived time series agree in magnitude. The HR time series is positively correlated with the AVISO counterpart because its representation of the Loop Current follows closely the observed in 2011. The LR run, on the other hand, does not capture the formation of a Ring in 2010 in late 2010 or the observed strengthening —-- without shedding --- of the Loop Current while into the northern Gulf in the summer of 2011, but is characterized by the detachment of a small-sized Ring in spring of 2011 that then propagates westward.

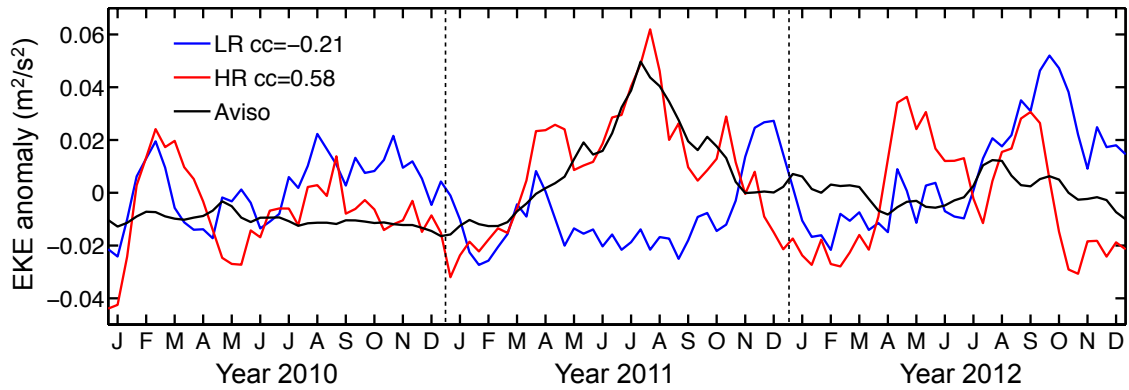

**Figure A1** Time series of mean EKE anomalies in the nested region [96.31ºW - 86.9ºW] and [25.40ºN - 30.66ºN] in the HR and LR simulations and in Aviso.

Modeled and in-situ temperature and salinity profiles over the period of 2010-2012 are compared in Fig. A2. Observed profiles (black lines) are obtained from CTD (Conductivity, Temperature, and Depth) measurements collected between August 2010 and June 2012 over 3 cruises (*R/V Oceanus*, OC468, August 22 – September 15, 2010; *R/V Endeavor*, EN496, July 3 – July 26, 2011; and *R/V Endeavor*, EN509, May 19 – June 19, 2012). Modeled profiles include the LR and HR simulations (blue and red lines, respectively) and the HYCOM-NCODA Gulf of Mexico 1/25º data-assimilative hindcast (GOMl0.04/expt_31.0, available at http://hycom.org/data/goml0pt04/expt-31pt0). The averages of 20 temperature and salinity profiles and plotted together with the model root-mean-square (RMS) error, along with two single profiles. Overall, ROMS, without any data assimilation, provides a reliable representation of the temperature and salinity distributions across the water column in the northern GoM in summer.

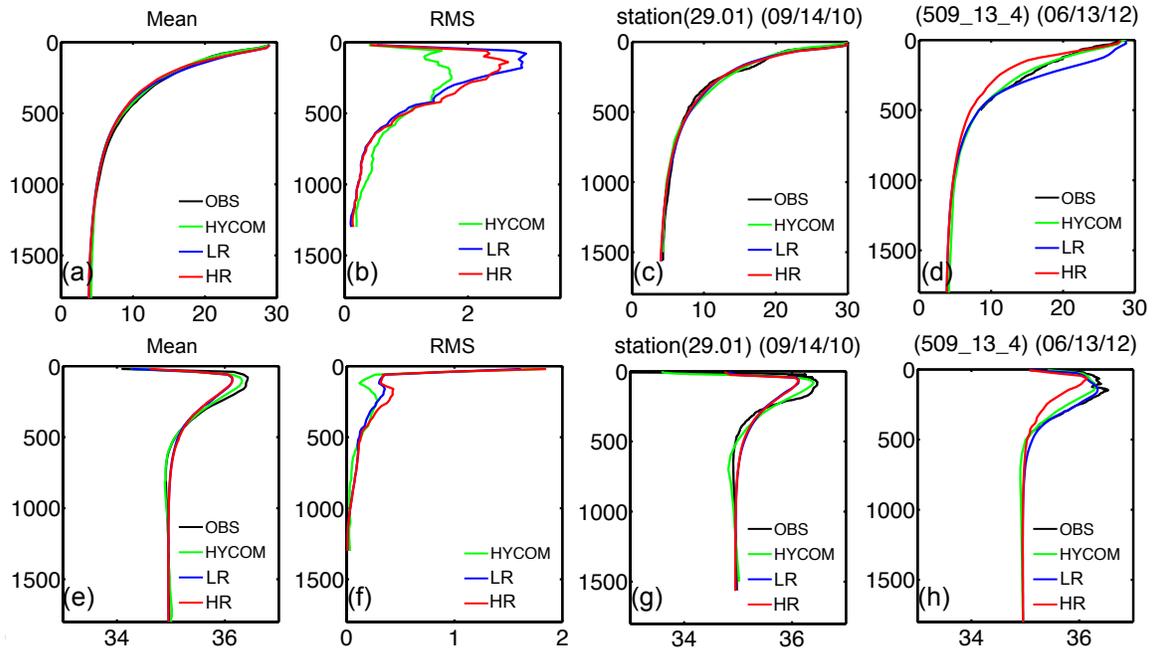

**Figure A2** Model and in situ profiles comparison: temperature (top row) and salinity (bottom row). In-situ profiles are in black, HYCOM results in green, and HR (LR) model outputs are in red (blue). Panel (a) (Panel (e)): mean temperature (salinity) profile computed using 20 in-situ observed profile collected during the Cruises OC468, EN496 and EN509. Panel (b) (Panel (f)): Root-Mean Square error between model outputs and in-situ profiles of temperature (salinity). Panel (c) (Panel (g)): in-situ temperature (salinity) profile collected in Sept. 2010 during Cruise OC468 at (87.93°W, 28.88°N). Panel (d) (Panel (h)): in situ temperature (salinity) profile collected in June 2012 during Cruise EN509 at (87.94°W, 27.43°N).

**Figure Captions**

**Figure 1** Model domain and bathymetry. The zoomed region is the nested area where the horizontal resolution is increased to 1.6 km.

**Figure 2** Modeled surface mean velocities over 2010-2012 superimposed on the mean speed, $\sqrt{\langle u^2 \rangle + \langle v^2 \rangle}$, in LR (left) and HR (right). Insets show the detail of the nested area where the two runs differ in horizontal resolution.

**Figure 3** Snapshots of surface salinity in HR (top) and HR$_{NOFW}$ (bottom) in winter (left) and summer (right) 2010. Unit: PSU

**Figure 4** Snapshots of normalized relative vorticity, $\zeta/f$, in the HR integration on February (Panel (a)), May (Panel (b)), August (Panel (c)), and November (Panel (d)) 15, 2011, and in the LR integration for winter (02/15/2011) and summer (08/15/2011) (Panels (e - f)).

**Figure 5** Annual cycle calculated over three years of (a) the local Rossby number, $R_o = |\zeta/f|$, at the ocean surface averaged over the nested domain in HR (red) and in LR (blue), and (b) of mixed layer depth (MLD) defined using a temperature criterion. The yearly evolution of $R_o$ in HR is shown for the whole region (solid thick lines), over areas where the water column is > 200 m (solid thin lines) and < 200 m (dashed). The averaged MLD is calculated over areas where the water column is > 200 m.

**Figure 6** Mean mixed layer depth calculated using a temperature criterion during the summer season (JAS) in (a) HR and (b) LR. Units: m.

**Figure 7** Seasonal PDFs of | $\zeta/f$ | in (a) HR and (b) LR normalized to have unit integral probability.

**Figure 8** Snapshots of normalized relative vorticity, |$\zeta/f$|, in February (a-c) and August (b-d) 2010 in the HR (left) and HR$_{NOFW}$ (right) runs.

**Figure 9** Time series of |$\zeta/f$| (red) and EKE (blue) from January 1st, 2010 to December 31st, 2010 in the HR$_{NOWF}$ (solid lines) and HR (dashed lines) simulations.

**Figure 10** Difference in mixed layer depth MLD during February 2010 between HR$_{NOFW}$ and HR. Unit: m. Only areas where the water column is > 200 m are shown; the domain average is 6.8 m.

**Figure 11** Top: mean L$_{D,ML}$ calculated for the winter (JFM) and summer (JAS) seasons over three years. Bottom: snapshots of L$_{D,MD}$ in February and August 15, 2011. The minimum value of L$_{D,MD}$ attained in August 15 is 4.7 km and is located to the east of the Mississippi mouth. Regions where the mixed layer reaches the bottom (white in figure) are excluded. Unit: m.

**Figure 12** Snapshots of vertical velocity *w* calculated at 4 m depth in the HR integration on February (Panel (a)), May (Panel (b)), August (Panel (c)), and November (Panel (d)) 15, 2011. Unit: $10^{-3}$ m s$^{-1}$.

**Figure 13** Annual cycle calculated at 4 m depth and using the three year long simulation of (a) | w |, (b) $\left|\left(\frac{\partial u}{\partial x}+\frac{\partial v}{\partial y}\right)/f\right|$, and (c) strain *S/f* (see text) averaged over the nested domain in HR (red) and in LR (blue) in the whole nested region (thick solid line), in HR over areas where the water column is < 200 m (dashed), and > 200 m (thin solid). Unit of *w*: $10^{-3}$ m s$^{-1}$.

**Figure 14** Snapshots of surface horizontal density gradients, $|\nabla_h \rho|$, in the HR integration in February (Panel (a)), May (Panel (b)), August (Panel (c)), and November (Panel (d)) 15, 2011, and in the LR integration for winter (02/15/2011) and summer (08/15/2011) (Panels (e - f)). Unit: $10^{-5}$ kg m$^{-4}$.

**Figure 15** Annual cycle of $|\nabla_h \rho|$ averaged areas where the water column is (a) < 200 m and (b) > 200 m over the nested domain in HR (red), LR (blue) and HR$_{NOFW}$ (black). Unit: kg m$^{-4}$.

**Figure 16** Snapshots of frontal tendency *F* in (a) February, and (b) August 2010 in HR (top) and HR$_{NOFW}$ (bottom). Unit: $10^{-14}$ kg$^2$ m$^{-8}$ s$^{-1}$.

**Figure 17** Annual cycle of frontal tendency $F$ in HR averaged over the nested domain and over three years in the whole region (thick solid line), over areas where the water column is < 200 m (dashed), and > 200 m (thin solid) and in $HR_{NOFW}$ over areas where the water column is > 200 m (black solid line). Unit: $kg^2\ m^{-8}\ s^{-1}$

**Figure 18** Annual cycle averaged over (a) 2010-2012 in HR and (b) 2010 in $HR_{NOFW}$ and over the nested domain of $PK$ (blue) and of $\langle |\nabla \bar{b}|\rangle^2_{xyz} \cdot \langle MLD \rangle^2_{xy}$. Units: $m^2 s^{-3}$ for $PK$ and $m^2 s^{-4}$ for $\langle |\nabla \bar{b}|\rangle^2_{xyz} \cdot \langle MLD \rangle^2_{xy}$.

**Figure 19** Annual cycle of $|\nabla_h \rho|$ in offshore waters separated in its mesoscale ($|\nabla_h \bar{\rho}|$, thin solid line) and submesoscale ($|\nabla_h \rho'|$, dashed line) components in HR averaged over three years and $HR_{NOFW}$ for 2010.

**Figure A1** Time series of mean EKE anomalies in the nested region [96.31°W - 86.9°W] and [25.40°N - 30.66°N] in the HR and LR simulations and in Aviso.

**Figure A2** Model and in situ profiles comparison: temperature (top row) and salinity (bottom row). In-situ profiles are in black, HYCOM results in green, and HR (LR) model outputs are in red (blue). Panel (a) (Panel (e)): mean temperature (salinity) profile computed using 20 in-situ observed profile collected during the Cruises OC468, EN496 and EN509. Panel (b) (Panel (f)): Root-Mean Square error between model outputs and in-situ profiles of temperature (salinity). Panel (c) (Panel (g)): in-situ temperature (salinity)

profile collected in Sept. 2010 during Cruise OC468 at (87.93ºW, 28.88ºN). Panel (d) (Panel (h)): in situ temperature (salinity) profile collected in June 2012 during Cruise EN509 at (87.94ºW, 27.43ºN).